\documentstyle[prl,aps,epsf,twocolumn]{revtex}

\begin{document}

\draft
\def\d{{\rm d}}
\def\LFP{L_{\rm FP}}

\title{
Transforming Gaussian diffusion into fractional,
a generalized law of large numbers approach 
}

\author{
E. Barkai}
\address{
Department of Chemistry
and Center for Materials Science and Engineering.
Massachusetts Institute of Technology.
77 Massachusetts Avenue
Cambridge, MA 02139.}
\date{\today}
\maketitle

\begin{abstract}

 The fractional Fokker--Planck equation (FFPE)
[R. Metzler, E. Barkai, J. Klafter  
{\em Phys. Rev. Lett.} 
{\bf 82},
3563 (1999)] describes an anomalous sub diffusive behavior of a particle
in an external force field. In this paper we present the solution
of the FFPE in terms of an integral transformation. The transformation
maps the solution of ordinary Fokker--Planck equation 
onto the solution of the FFPE. We investigate in
detail the force free particle and the particle in uniform
and harmonic fields.
The meaning of the transformation is explained based on the asymptotic
solution of the continuous time random walk (CTRW).
We also find an exact solution of the CTRW and compare the CTRW
result with the integral solution of the FFPE for the
force free case. 

\end{abstract}

\pacs{02.50.-a,r05.40.Fb,05.30.Pr}

%

\section{Introduction}

 The continuous time random walk (CTRW),
introduced by Montroll and Weiss \cite{Weiss7,Weiss1},
describes different types of diffusion processes including the standard
Gaussian diffusion, sub diffusion 
\cite{Weiss1,shlesinger,SM,Bouch} and
L\'evy walks \cite{KBS,SKW,Klafter1,barkai8,barkai10,Antony,latora}.
The CTRW has been a useful tool in many diverse
fields and over the last three decades
e.g., it was used to describe transport in disorder medium
\cite{SM,Silbey,Levitz} and in low dimensional
chaotic systems \cite{Klafter1,Zumofen,Swin,barkai11}.
More recently fractional kinetic equations were investigated
as a tool describing phenomenologically anomalous diffusion
\cite{kuz,fogedby1,zaslavsky,Mainardi,Kol,Rocco,Hui}.
In these equations fractional derivatives \cite{oldham}
replace ordinary integer
derivatives in the standard integer kinetic equation.
It is believed that fractional calculus can be used to model
non integer diffusion phenomena.
 Schneider and
Wyss have formulated the fractional diffusion equation
(see Sec. \ref{secFDE}) \cite{schneider} and
under certain conditions \cite{Hilfer,compte,barkai9}
the  fractional diffusion equation describes 
the asymptotic large time behavior of the decoupled
sub-diffusive CTRW (with $\langle  r^2 \rangle \sim t^{\delta}$; $\delta<1$).
Thus some what like ordinary random walks which can be described
asymptoticly by the diffusion equation, so can the 
CTRW be described by fractional diffusion equation.

 Recently, Metzler, Barkai and Klafter \cite{MBK} 
have investigated fractional Fokker--Planck  
equation (FFPE)
to describe an anomalous sub-diffusion motion
in an external non--linear force field. 
In the absence of the external force field the FFPE reduces
to the fractional diffusion equation.
The FFPE is 
an asymptotic equation which 
extends the CTRW to include  the effect of an external force field.
The CTRW itself was  not designed for a description of a particle
in external force field thus the FFPE
describes new type of behavior not explored in depth yet.  

 The main result in this paper concerns a transformation
of ordinary Gaussian diffusion into fractional diffusion 
\cite{Bouch,Gurt,saichev,BS}. 
Let $P_{\alpha}({\bf r},t) $ be
the propagator (Green's function) of
fractional Fokker--Planck equation 
(a special case is the fractional diffusion
equation),
here $0<\alpha <1$ is the fractional exponent and $\alpha=1$ is the
standard case described by ordinary Fokker-Planck equation. 
Then the solution, $P_{\alpha}({\bf r},t)$,
is found based on the transformation 
\begin{equation}
P_{\alpha}({\bf r},t)=\int_0^{\infty}
\left[{d  N(s,t)\over d s}\right] P_1 ({\bf r},s) ds,
\label{eq111}
\end{equation}
and 
\begin{equation}
N(s,t)=1-L_{\alpha}\left({t\over s^{1/\alpha}}\right)
\label{eq411}
\end{equation}
denotes the ``inverse'' one sided L\'evy stable distribution
(i.e., $L_{\alpha} (x)$ is the one sided L\'evy stable distribution). 
For example  consider the force free case.
The solution of the FFPE 
$P_{\alpha}({\bf r}, t)$, with initial conditions
concentrated on the origin, 
is found by transforming 
$P_1({\bf r}, s)$
and the transformed function is 
the well known Gaussian solution of  
the integer diffusion equation with initial conditions 
concentrated on the origin.
The transformation Eq. (\ref{eq111}) besides its practical
value for finding solutions of  
the FFPE also explains its
meaning and relation to the CTRW (see details below).

We list previous work on the integral transformation Eq. (\ref{eq111})\\
{\bf 1}  
Bouchaud and  Georges
\cite{Bouch}
for the asymptotic long time limit of the CTRW, 
then $P_1({\bf r}, s )$ is Gaussian,\\ 
{\bf 2}  Zumofen and Klafter \cite{Gurt} for
CTRW on a lattice, then $P_1({\bf r},s)$
is solution of an ordinary random walk on a lattice,\\ 
{\bf 3} Saichev and Zaslavsky 
\cite{saichev} 
for one dimensional solution of fractional diffusion equation,
they have also considered an extension which includes L\'evy flights,\\ 
{\bf 4} 
Barkai and Silbey \cite{BS} 
for the fractional Ornstein-Uhlenbeck process.\\
We generalize these results for the dynamics described by the FFPE,
and investigate in greater detail Eq. 
(\ref{eq111}) in the context of CTRW
and fractional diffusion equation in dimensions $d=1,2,3$.

 This paper is organized as follows. 
In Sec. (\ref{secCTRW}) we follow \cite{Bouch}
and  derive the transformation
Eq. (\ref{eq111}) based on the CTRW.
We find an exact solution to the CTRW valid for short
and long times and compare this solution with the result obtained
from Eq. (\ref{eq111}). 
In Sec. (\ref{secFDE}) we consider the fractional diffusion equation.
We investigate its integral solution Eq. (\ref{eq111}) as well
as the Fox function solution found in 
\cite{schneider}.
In Sec. (\ref{secFFPE}) we show how to obtain solutions
of the FFPE using Eq. (\ref{eq111}). We give as examples
the motion of a particle in a uniform and
harmonic force fields. In Sec. 
(\ref{secSUM}) a brief summary is given.

\section{Continuous Time Random Walk}
\label{secCTRW}

 In the decoupled version of the continuous time random walk (CTRW),
a random walker hops from site to site and at each site it is trapped
for a random time \cite{Weiss1}. For this well known model two 
independent probability
densities describe the random walk. The first is $\psi(t)$, the probability
density  function (PDF) of  the independent identically distributed (IID)
pausing times between successive steps.
The second is the PDF $f\left( r \right)$ for the IID displacements
of the random walker at each step. 
Thus the CTRW describes a  process for which the particle
is trapped on the origin for time $\tau_1$, it then jumps
to ${\bf r_1}$, it is trapped on ${\bf r_1}$ for time $\tau_2$
and then it jumps to a new location, the process is then renewed.
In what follows we assume $f(r)$ has a finite variance and a zero mean.
The asymptotic behavior of the 
 decoupled CTRW  is well investigated 
\cite{Weiss1,shlesinger,Gurt,Tunaley,BHW,WWH}
and we now summarize some results from the CTRW literature
which are of relevance to our work.

 Let $P\left(r, t\right)$
be the PDF of finding the CTRW particle at $r$ at time $t$. 
Let $N_{CT}\left(s, t \right)$ be the probability that $s$ steps are
made in the time interval $( 0 , t)$ so clearly
\begin{equation}
\sum_{s=0}^{\infty} N_{CT}(s,t) =1
\label{eqAmm1}
\end{equation}
and the subscript CT denoted the CTRW.
Because the model is decoupled 
\begin{equation}
P\left( r, t\right)=\sum_{s = 0}^{\infty} N_{CT}(s,t) W\left(r,s\right),
\label{eq01}
\end{equation}
and $W \left(r,s\right)$ 
is the probability density that a particle has 
reached $r$ after $s$ steps. In what follows we shall use
the Fourier $({\bf r} \to {\bf k})$ and Laplace
$(t \to u)$ transforms, we use the
convention that the argument of a function indicates 
in which space the function is defined, e.g., 
\begin{equation}
P\left( r, u\right)=\sum_{s = 0}^{\infty} N_{CT}(s,u) W\left(r, s\right),
\label{eq01asd}
\end{equation}
is the Laplace transform of $P(r, t )$.

$W \left( r,s \right)$ will generally depend on $f(r)$, however
we are interested only in the large time behavior of $P\left( r, t\right)$
meaning that only contributions from large $s$ are important.
From standard central limit theorem we know
\begin{equation}
W\left( r,s \right) \to_{s \to \infty} G\left( r,s \right)= 
{ 1 \over \left(4 \pi s\right)^{d/2}} \exp\left( - {r^2 \over 4 s} \right).
\label{eq02}
\end{equation}
We have used the assumption that the system is unbiased and 
use convenient units.
For most systems and for large times $N_{CT}(s,t)$ is concentrated
on $s=t / \tau_{av}$ and $\tau_{av}=\int_0^{\infty} t \psi(t) dt$ is the averaged
pausing time. In this case 
$P\left( r, t\right)$ will become Gaussian when $t \to \infty$.
When $N_{CT}\left(s,t\right)$ is broad
a non-Gaussian behavior is found.
This is the case when $\tau_{av}=\infty$ or in other words
when the Laplace transform of $\psi(t)$; $\psi(u)$ behaves as
\begin{equation}
\psi\left( u \right) \sim 1 -  u^{\alpha} + \cdots \ \ 0< \alpha \le 1,
\label{eq03}
\end{equation}
and we have used convenient units.
Shlesinger \cite{shlesinger} showed that in this case an anomalous sub-diffusive behavior is found, 
$\langle r^2 \rangle \sim t^{\alpha}$. 
Because the steps are independent and based on convolution theorem of Laplace transform
$$ N_{CT} \left(s, u \right) = $$
$$ { 1 - \psi\left( u \right) \over u} 
\exp\left\{ s \ln \left[ \psi\left(u\right)\right] \right\} \sim  $$
\begin{equation}
u^{ \alpha - 1} \exp\left( - s  u^{\alpha} \right),
\label{eq04}
\end{equation}
and $N_{CT}(s,u)$ is the Laplace transform of $N_{CT}(s,t)$. 

 Following 
\cite{Bouch}
we replace the summation in
Eq. (\ref{eq01asd}) with integration 
and use 
Eq. (\ref{eq02}) to find
\begin{equation}
P(r, u )\sim \int_0^{\infty}  n(s,u) G(r,s) ds
\end{equation}
and 
according to Eq. (\ref{eq04}) 
\begin{equation}
n(s,u) \equiv  u^{ \alpha - 1} \exp\left( - s  u^{\alpha} \right).
\label{eq04aa}
\end{equation}
  Using the inverse Laplace transform
we find
\begin{equation}
P\left( r, t\right)\sim \int_{0}^{\infty} n(s,t) G\left( r,s\right)ds
\label{eq05}
\end{equation}
and $n(s,t)$ is the inverse Laplace transform of
$n(s,u)$ \cite{Bouch,saichev,BS}
\begin{equation}
n(s,t) =
{ 1 \over \alpha}   {t\over s^{ 1 + 1/\alpha}  }  l_{\alpha} \left( { t \over s^{ 1 /\alpha} } \right)
\label{eq06}
\end{equation} 
 $l_{\alpha}\left( z \right)$ in Eq. (\ref{eq06})
is a one sided L\'evy stable probability  density
whose Laplace transform is
\begin{equation}
l_{\alpha}\left( u \right) = \int_0^{\infty} e^{ - u x} 
l_{\alpha}\left( x \right) dx=e^{-u^{\alpha}}.
\label{eq06b}
\end{equation}

 According to Eq. (\ref{eq05}) the large time behavior of the CTRW solution
is reached using an integral transformation of the Gaussian  solution
of ordinary diffusion processes [i.e., of $G(r,s)$]. 
As we shall see
similar transformations can be used to solve the FFPE and in particular the
fractional diffusion equation. 

  The kernel $n \left(s, t\right)$ is a non negative PDF
normalized according to
\begin{equation}
\int_0^{\infty} n\left(s, t \right) ds=1,
\label{eq06ba}
\end{equation}
it replaces the CTRW probability $N_{CT}(s,t)$.
Notice that the PDF
$n\left(s, t\right)$, like $N_{CT}(s,t)$, 
is independent of the dimensionality of the problem
$d$. Some properties of $n(s,t)$ are given in \cite{saichev}.
All moments of $n(s,t)$ are finite and are
given later in Eq.
(\ref{eqB06}).
It is easy to show that $n(s,t)=[d N(s,t) / ds]$ with $N(s,t)$
defined in Eq. 
(\ref{eq411}). $n(s,t)$ is called the ``inverse''
one sided L\'evy stable probability density \cite{Marcin}.

 When $\alpha = 1$, we find $n(s,t)=\delta(s- t)$
and hence $P(r,t)$ is Gaussian. This is 
expected since when $\alpha= 1$ the first moment
of pausing times $\tau_{av}$ is finite, therefore the law of large numbers
is valid, and hence we expect that the number of steps
$s$ in the random walk scheme will follow $s \sim t/\tau_{av}$.
When $\alpha< 1$ the law of large numbers
is not valid and instead the random number of steps
$s$ is described by $n(s,t)$. Thus the transformation, Eq. (\ref{eq05}),
has a meaning of a generalized law
of large numbers.

Eq. (\ref{eq06}) gives the kernel $n(s,t)$ in terms of
a one sided L\'evy stable density.
Schneider \cite{Sch1} has expressed L\'evy stable densities
in terms of a Fox H function \cite{sri,mathai}
\begin{equation}
l_{\alpha}(z) = { 1 \over \alpha z^2} H_{1,1}^{1,0}\left[ z^{-1} \left|\right.
\begin{array}{c}
\left( - 1, 1 \right) \\
\left( - 1/ \alpha, 1 / \alpha\right) 
\end{array}
\right]. 
\label{eqLeSc}
\end{equation}
Asymptotic behaviors of the one sided L\'evy density can be found
in Feller's book \cite{Feller} or based on the asymptotic behaviors
of the H Fox function \cite{Sch1,sri,mathai}. 
For the cases $\alpha=1/3,1/2,2/3$, 
closed
form equations in terms
of known functions, may be found in
\cite{Sch1} (notice that \cite{Sch1} points out relevant errors in the
literature on
L\'evy stable densities).

Important for our purposes is the result obtained by Tunaley \cite{Tunaley}
already in $1974$ which expressed the asymptotic behavior of the CTRW 
solution $P\left( r, t \right)$,
in terms of its Fourier--Laplace transform, as shown  in \cite{Tunaley}
\begin{equation}
P\left({\bf k}, u \right)  \sim  { u^{\alpha-1} \over u^{\alpha} + {\bf k}^2}  .
\label{eqa02}
\end{equation}
In Appendix A we verify that Eq. (\ref{eqa02}) is indeed the Fourier--Laplace
transform of Eq. (\ref{eq05}). The inversion 
of equation (\ref{eqa02}) was accomplished by
Tunaley \cite{Tunaley} in one dimension
and by Schneider and Wyss \cite{schneider} in dimensions
two and three (see more details below). 

\subsection{CTRW Solution}

\begin{figure}[htb]
\epsfxsize=20\baselineskip
\centerline{\vbox{
      \epsffile{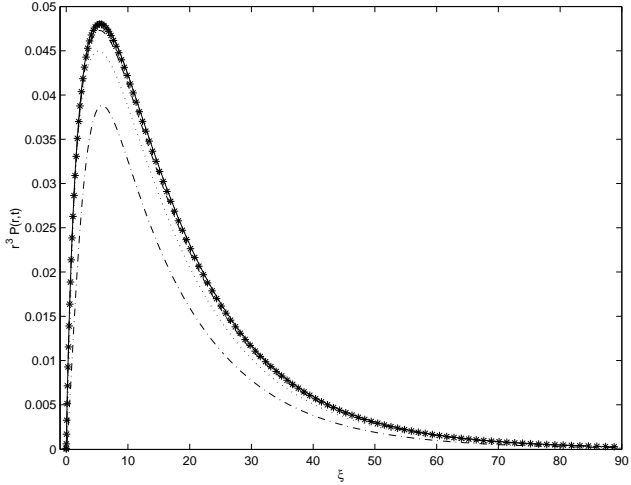}  }}
\caption {
We show $r^3 P(r,t)$ versus $\xi=r^2/t^{\alpha}$
for the CTRW process and for $\alpha=1/2$,
$d=3$. 
The curves in the figure, $t=2$ (dot dashed), $t=20$ (dotted),
$t=200$ (dashed) and $t=2000$ (stars) are converging in the limit 
$t \to \infty$ to the asymptotic master curve (solid).
This master curve was computed numerically based on 
Eq.
(\protect{\ref{eq05}}).
It is difficult to distinguish between the results for $t=200,2000$ 
and the asymptotic result.
Not shown is the delta function contribution
at $r=0$. 
}
\label{figone}
\end{figure}

 Let us consider an example of the (decoupled) CTRW process.
The solution of the CTRW in ${\bf k, u }$ space is \cite{Weiss1}
\begin{equation}
P\left( {\bf k } , u \right) = { 1 - \psi(u) \over u} { 1 \over 1 - \psi(u)f(\bf{k})  }.
\label{eqctrwks}
\end{equation}
Usually CTRW
solutions are found based on numerical inverse Fourier--Laplace transform
of Eq. (\ref{eqctrwks}). 
Here we find an exact solution of the CTRW process for a special choice
of $f(r)$ and $\psi(t)$. Our solution is an infinite sum of well known
functions.  We assume
the PDF of jump times $\psi(t)$ to be one sided L\'evy stable
density with $\psi(u)=\exp(-u^{\alpha})$ 
\cite{remark}. Displacements are
assumed to be Gaussian and then
\begin{equation}
W \left(r,s \right) = 
{ 1 \over \left( 4 \pi s\right)^{d/2}} \exp\left( - {r^2 \over 4 s} \right).
\label{eq02aaa}
\end{equation}
is exact, not only asymptotic. For this choice of PDFs
the solution of the CTRW can be found explicitly.
We use 
\begin{equation}
N_{CT}(s,u)={1 - \exp(-u^{\alpha})\over u} \exp\left( - s u^{\alpha} \right)
\label{eqeexx}
\end{equation}
and the convolution theorem of Laplace transform
to find
$$ N_0(t)=1-L_{\alpha}\left(t\right)$$
\begin{equation}
N_{CT}(s,t) = L_{\alpha}\left({t\over s^{1/\alpha}}\right)
-L_{\alpha}\left({t\over (s+1)^{1/\alpha}}\right)
\label{eqadhjf}
\end{equation}
and 
\begin{equation}
L_{\alpha}\left(t\right) = \int_0^{t} l_{\alpha}(t) dt 
\label{eqhjf}
\end{equation}
is the one sided L\'evy stable distribution.
Inserting Eqs. (\ref{eq02aaa}), (\ref{eqadhjf})
in Eq.
(\ref{eq01}) we find
$$P(r,t) = \left[ 1 - L_{\alpha}\left( t\right) \right]\delta\left(r\right)+$$
$$ \sum_{s=1}^{\infty}\left\{ L_{\alpha}\left[ {t\over s^{1/\alpha} }\right]
-L_{\alpha}\left[  {t \over\left(s+1\right)^{1/\alpha}} \right] \right\} 
\times $$
\begin{equation}
{ 1 \over \left(4\pi s\right)^{d/2}} \exp\left( - {r^2 \over 4 s} \right).
\label{eqexct}
\end{equation}
The first term on the right hand side describes random walks 
for which the particle did not leave the origin within
the observation time $t$, the other terms describe random
walks where the number of steps is $s$.
In Fig. \ref{figone} we show the solution of the CTRW
process in a scaling form. We consider a three dimensional
case, $\alpha = 1/2$ and use
\begin{equation}
L_{1/2} \left( t \right) =
{ 1 \over \sqrt{\pi} } \Gamma\left[1/2,1/(4 t) \right].
\label{eqgammm}
\end{equation}
Here $\Gamma(\cdot,\cdot)$ denotes the incomplete Gamma function.
The figure shows $r^3 P(r,t)$ versus the scaling variable
$\xi=r^2/t^{\alpha}$. For large times $t$ the solution converges to
the asymptotic solution found based on the integral transformation
Eq. (\ref{eq05}).


%
%

 Let us calculate the Cartesian moments
$$ M\left( 2 m_1 , \cdots, 2 m_d\right)=  $$
\begin{equation}
\int_{-\infty}^{\infty} dx_1
\cdots\int_{-\infty}^{\infty} dx_d x_1^{ 2 m_1} \cdots x_d^{ 2 m_d} 
P\left(x_1,\cdots,x_d, t \right)
\label{eqIS06}
\end{equation}
with non negative integers $m_1,m_2, \cdots$.
Clearly the odd
moments are equal zero and from normalization
$M(0,\cdots,0)=1$. In Appendix B we find
$$ M\left( 2 m_1 , \cdots, 2 m_d\right)=  $$
\begin{equation}
C_{m,d}\sum_{s=1}^{\infty} \left\{ L_{\alpha}\left[ { t \over s^{1 /\alpha}}\right]-
L_{\alpha}\left[ { t \over \left( s + 1\right)^{ 1 / \alpha}}\right]\right\}s^m
\label{eqffff}
\end{equation}
with
\begin{equation}
C_{m,d}={ 2^{2 m} \over \pi^{d/2}}\Pi_{i=1}^d \Gamma\left( m_i + {1 \over 2} \right) 
\label{eqffff1}
\end{equation}
and $m=\sum_i^d m_i> 0$. In the Appendix we also show
that for $t \to \infty$ 
$$ M\left( 2 m_1 , \cdots, 2 m_d\right)\sim  $$
\begin{equation}
C_{m,d} \Gamma(1+m) 
{ t^{\alpha m} \over \Gamma\left(1 + \alpha m\right)}.
\label{eqddd}
\end{equation}
To derive Eq. (\ref{eqddd}) we used 
the small $u$ expansion of the Laplace transform of Eq. (\ref{eqffff})
and Tauberian
theorem. 
Later we shall show that the moments 
in Eq. (\ref{eqddd}) are identical to the moments
obtained directly from the integral transformation Eq. (\ref{eq05}). 
Hence
our interpretation of Eq. (\ref{eq05}) as the
asymptotic solution of the CTRW is justified, 
[however our derivation of Eq. (\ref{eqddd}), is based
on a specific choice of $\psi(t)$ and $f(r)$]. 

\section{Fractional Diffusion Equation}
\label{secFDE}

 The fractional diffusion equation describes the asymptotic behavior of the
CTRW Eq. (\ref{eq05}). 
Balakrishnan \cite{Balakrishnan} has derived a fractional diffusion process
based upon a generalization of Brownian motion in one dimension.
Schneider and Wyss \cite{schneider,Wyss} have formulated the 
following fractional
diffusion equation describing this process
\begin{equation}
{\partial P\left(r, t \right) \over \partial t} =  _0 D_t^{1 - \alpha} \nabla^2 P\left( r, t \right),
\label{eqa01}
\end{equation}
and $0< \alpha < 1$.
The fractional Riemann Lioville derivative 
$_0 D_t^{1 - \alpha}$ in Eq. (\ref{eqa01}) is  defined  \cite{oldham}
\begin{equation}
_0D^{1- \alpha}_t Z\left(t\right) = { 1 \over \Gamma(\alpha)} {\partial \over \partial t}
\int_0^t dt' { Z(t') \over (t- t')^{1 - \alpha}}.
\label{eq20aa}
\end{equation}
The Fourier--Laplace 
solution of the fractional diffusion equation, with initial conditions 
$P({\bf r},t=0)=\delta({\bf r})$,
\begin{equation}
P\left({\bf k}, u \right)  =  { u^{\alpha-1} \over u^{\alpha} + {\bf k}^2}  
\label{eqa02a}
\end{equation}
is identical to the right hand side of Eq. (\ref{eqa02}).
Therefore
the integral solution of fractional diffusion
equation is
\begin{equation}
P\left( r, t\right)= \int_{0}^{\infty} n(s,t) G\left(r,s\right)ds.
\label{eq05b}
\end{equation}
Thus solution Eq. (\ref{eq05}) which was only asymptotic in the CTRW
framework is exact within the fractional diffusion equation approach.
The integral solution Eq. (\ref{eq05b}) for $d=1$ was investigated
by Saichev and Zaslavsky, we investigate also the cases $d=2,3$
which exhibit behaviors different than the one dimensional case.

  As mentioned, Schneider and Wyss \cite{schneider} 
have inverted Eq. (\ref{eqa02a}) finding the solution to the fractional 
diffusion equation in terms of rather formal Fox
function. 
Later we shall show that moments of $P(r,t)$ calculated
based on the integral solution Eq. (\ref{eq05b}) in dimensions $d=1,2,3$ are
identical to the moments calculated based upon the Fox function solution.
In this sense we show that the two solutions are identical.

 At this stage we have only shown that the two approaches, asymptotic CTRW
and fractional diffusion equation are identical.  The reader might be wondering
why should we bother with the fractional equation if the CTRW approach gives
identical results. The situation is some what similar 
to the standard diffusion equation
which predicts a Gaussian evolution.  The diffusion equation $\alpha = 1$ is
an asymptotic equation   
describing much more general
random walks. The main advantage of the diffusion equation
over a random walk approach is its simplicity. Solutions with special boundary  
conditions (reflecting/absorbing) are relatively simple and usually 
capture the essence
of the more complex random walks. Another extension of the diffusion equation
is the diffusion in external field as described by Fokker--Planck equations,
such an extension for random walks is cumbersome. 
The same is true for the fractional
diffusion equation. It can serve as a phenomenological tool
describing anomalous diffusion. As we shall
show in the next section we can include the effects of an external
field.  On the other hand CTRW  by definition is not
built to consider an external field.


\subsection{Integral Solution}

 We first notice that the integral solution Eq.
(\ref{eq05b}) shows that $P\left(r, t \right)$ is normalized
and non-negative. The normalization is easily seen with the
help of Eq.
(\ref{eq06ba})
and the non negativity of $P\left( r, t \right)$ is
evident because both $n(s,t)$ and $G(r,s)$ are non
negative.

 Calculation of spherical moments $\langle r^n(t) \rangle$ is now
considered. The calculation follows two steps, the first is to calculate
$\langle r^n (s ) \rangle$ for Gaussian diffusion
(or find the Gaussian moments in a text book) then using the
transformation defined in  Eq. 
(\ref{eq05b}) we find the moments $\langle r^n (t) \rangle$ for the fractional
case. More precisely in Laplace $u$ space
\begin{equation}
\langle r^n\left( u \right) \rangle = \int_0^{\infty} ds n(s, u )   
\left[ \int G\left( r,s \right) r^n d {\bf r} \right],
\label{eqIS01a}
\end{equation}
the Gaussian spherical moments are
\begin{equation}
\int G\left( r,s \right) r^n d {\bf r} 
=\Omega_d \Gamma\left( { n + d \over 2} \right) 2^n s^{n/2}
\label{eqIS02}
\end{equation}
with $\Omega_1=1/\sqrt{\pi}$, $\Omega_2=1$ and $\Omega_3=2/\sqrt{\pi}$.
The moments for the fractional case are found using Eq.
(\ref{eqIS01a})
$$ \langle r^n\left( u \right) \rangle = $$
$$  \Omega_d 2^n \Gamma\left( { n + d \over 2 } \right)
\int_0^{\infty} ds s^{n/2} n(s,u)= $$
\begin{equation}
\Omega_d 2^n \Gamma\left( { n + d \over 2 } \right)\Gamma(1+n/2) u^{ - 1 -\alpha n/2}.
\label{eqIS03}
\end{equation}
Using the inverse Laplace transformation we find
\begin{equation}
\langle r^n \left( t \right) \rangle = \Omega_d { \Gamma\left( { n + d \over 2} \right) 2^n   
\Gamma\left( 1 + n/2\right) \over \Gamma\left( 1 + \alpha n / 2  \right) } t^{\alpha n / 2}.
\label{eqIS04}
\end{equation}
For $n=2$
\begin{equation}
\langle r^2 \left( t \right) \rangle = { 2 d t^{\alpha} \over \Gamma\left( 1 + \alpha \right) }.
\label{eqIS05}
\end{equation}
When $\alpha=1$ the moments in  Eq. (\ref{eqIS04}) are identical
to the Gaussian moments in Eq. (\ref{eqIS02}).

 In Appendix C we calculate the Cartesian moments defined in
Eq. (\ref{eqIS06}).
We find 
\begin{equation}
 M\left( 2 m_1 , \cdots, 2 m_d\right) = 
C_{m,d}
{\Gamma\left( 1 + m \right) \over \Gamma\left( 1 + \alpha m \right) }t^{ \alpha m } 
\label{eqIS07}
\end{equation}
which is identical to the right hand side of equation (\ref{eqddd}).
Thus the moments of the integral solution of fractional
diffusion equation are identical to the asymptotic behavior
of the moments of the CTRW found in the previous section. 

  Eq. (\ref{eqIS07}) was also found in \cite{schneider} based
on Fox function solution of fractional diffusion equation 
(see details next subsection).
Since the integral solution
Eq. (\ref{eq05b})
gives identical
moments to those found using the Fox function solution one can assume that the
two solutions are identical.
It would be interesting to show in a more direct way
that the Fox function solution is identical to the integral
solution. In Appendix D we use a theorem on Fox
functions and attempt to give such a proof. Unfortunately
we fail. The  interested reader 
is referred to Appendix D. 
 
\subsection{Fox Function Solution}

In  \cite{schneider}, the Mellin transform was used to find
the solution of the fractional diffusion equation in terms of Fox  H function
$$ P\left( r , t \right) = \alpha^{-1} \pi^{-d/2} r^{ - d} \times $$
\begin{equation}
H_{12}^{20}
\left( 2^{- 2 /\alpha} r^{2 /\alpha} t^{-1} |
\begin{array}{l}
\left( 1 , 1 \right) \\
\left( d/2, 1/\alpha),(1,1/\alpha\right)
\end{array}
\right)
\label{eq23}
\end{equation}
The asymptotic
expression for this solution is \cite{schneider}
\begin{equation}
 P\left( r , t \right) \sim \kappa^{\alpha} r^{-d} \xi^{{d \over 2 \left( 2 - \alpha
\right)}}\exp\left(- \lambda_1 \xi^{1 \over 2 - \alpha} \right),
\label{eq24}
\end{equation}
where $\xi\equiv r^2/t^{\alpha}$ is the scaling variable,
\begin{equation}
\kappa^{\alpha} = \pi^{- d/2} 2^{ - {d \over 2 - \alpha} } \left( 2 - \alpha
\right)^{-1/2}
 \alpha^{ \left[ \alpha\left( d + 1 \right) /2 - 1 \right]/\left( 2 -
\alpha) \right]}
\label{eq25}
\end{equation}
and
\begin{equation}
\lambda_1 = \left( 2 - \alpha \right)
\alpha^{\alpha/\left( 2 - \alpha\right)}2^{-2/\left( 2 - \alpha\right) }.
\label{eq26}
\end{equation}
Eq. (\ref{eq24}) is valid for $\xi >> 1$.
For a brief introduction to Fox functions and its application see 
\cite{Gl}.

 The behavior for $\xi << 1$ is found in Appendix E based
on the asymptotic expansion of the $H$ function \cite{sri,mathai}.
For dimensions
$d=1$
we find  
\begin{equation}
P\left( r , t \right) = { 1 \over 2 t^{\alpha/ 2} } \sum_{n = 0}^{\infty}
{ \left( - 1 \right)^n \xi^{ n/2} \over n!\Gamma\left[ 1 - \alpha(n+1)/2\right]}
\label{eqA14}
\end{equation}
and for $d=3$
\begin{equation}
P\left( r , t \right) = { 1 \over 4 \pi t^{3 \alpha/2} \xi^{1/2} }
\sum_{n=0}^{\infty} { \left( - 1 \right)^n \xi^{ n/2} \over n! 
\Gamma\left[ 1 - \alpha(1+n/2) \right] }.
\label{eqA15}
\end{equation}
The leading terms in these expansions  are for $d=1$
\begin{equation}
P\left( r , t \right)  = { 1 \over 2 \Gamma\left( 1 - \alpha/2\right)} 
{ 1 \over t^{\alpha/2}}
\cdots
\label{eqA12a}
\end{equation}
and for $d=3$
\begin{equation}
P(r,t) = { 1 \over 4 \pi \Gamma\left( 1 - \alpha\right)} 
{ r^{-1} \over t^{\alpha}} + \cdots .
\label{eqA13a}
\end{equation}
We see that for $d=1$ and $\alpha=1$, $P(r=0, t ) = 1/(2 \sqrt{\pi t})$,
as expected for this normal case.
We  also see that for $d=3$, $\alpha\ne 1$ and
when $r \to 0$ the solution diverges like
$P(r,t) \sim 1/r$. This behavior is not unphysical and $P(r,t)$ is normalized according
to $4 \pi \int P(r,t) r^2 dr = 1$.

For $d=2$ the asymptotic expansion of the Fox function is not known
(see more details in Appendix E).
Using Eq. (\ref{eq05b}) one can show that for $d=2$, and $\xi<<1$
\begin{equation}
P(r,t) \sim { 1 \over \pi \Gamma(1 - \alpha) t^{ \alpha} } 
\ln\left[t^{\alpha/2}/r\right].
\label{eqASDF}
\end{equation} 
Eqs. (\ref{eqA12a}-\ref{eqASDF}) were derived independently by A. I. Saichev
\cite{SaichevPC}.


The CTRW behavior on the origin is different than
what we have found for the fractional diffusion equation 
Eqs. (\ref{eqA12a}-\ref{eqASDF}).
Within the CTRW on the decay on the origin is described with  
the first term on the right hand side of Eq. (\ref{eqexct}),
$\left[ 1 - L_{\alpha}\left( t\right) \right]\delta\left(r\right)$.
This term describes
random walks for which the particle is trapped on its
initial location during the time of observation $t$.
This  CTRW term decays like $t^{-\alpha}\delta\left(r\right)$ for long times
and 
no such singular term is found in the solution of the fractional.
In dimensions $d=2,3$ the decay of the CTRW singular term is as slow
as the decay of $P(0,t)$ found from the fractional diffusion equation.
Hence on the origin the fractional diffusion approximation does not work
well. In contrast,
for normal random walks, the singular term
decays exponentially with time and then the diffusion approximation
works well already after an exponentially short time.

\subsection{Example}

Here we use the integral solution
of the fractional diffusion equation
for a specific example. The Fox function solution
are not tabulated and hence from a practical point
of view one has to 
use numerical methods to find the solution. 
Also the convergence of the asymptotic solution  for $\xi<<1$ 
is not expected to be fast and hence the integral solution is of special
practical use.

\begin{figure}[htb]
\epsfxsize=20\baselineskip
\centerline{\vbox{
      \epsffile{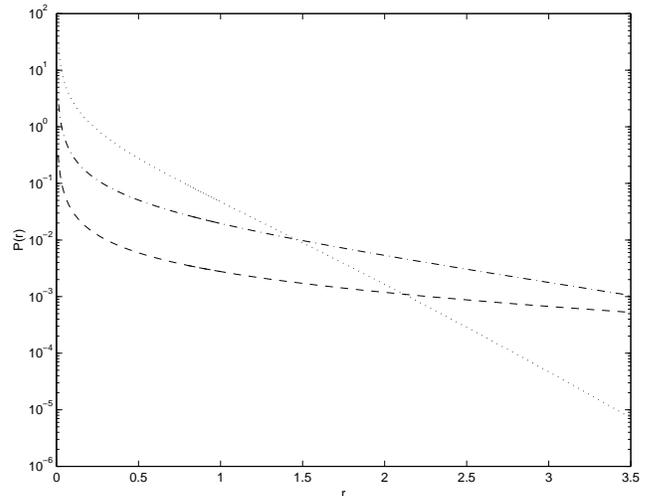}  }}
\caption {
$P(r,t)$ versus $r$ on a semi log plot and for different times 
$t=0.02$ (dotted), $t=2$ (dash dotted) and $t=200$  (dashed).
Note that  $P(r=0,t)=\infty$  hence the solution in figure is cutoff close to
$r\to 0$. 
}
\label{fig1}
\end{figure}

 We consider the case $\alpha = 1/2$ and $d=3$.
Using
\begin{equation}
l_{1/2}\left( z\right) = 
{ 1 \over 2 \sqrt{\pi} } z^{ - 3/2} \exp\left[ - 1/ \left(4 z\right) \right]
\label{eqExam}
\end{equation}
with $z> 0$. We find the  integral solution $P(r,t)$ Eq. (\ref{eq05b}) 
using numerical integration. We have used Mathematica which gave all
the numerical results without difficulty.
In Fig. \ref{fig1} we show $P(r,t)$ 
versus r on a semi log plot and for different times $t$.
Close to the origin $r=0$ we observe a sharp increase of $P(r,t)$ as
predicted in Eq. 
(\ref{eqA13a}).

\begin{figure}[htb]
\epsfxsize=20\baselineskip
\centerline{\vbox{
      \epsffile{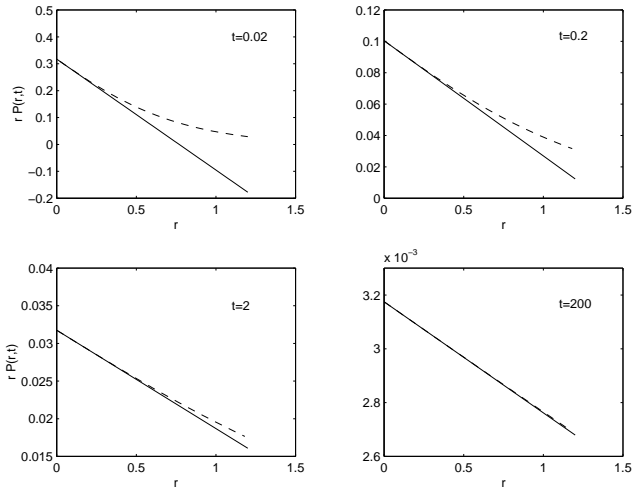}  }}
\caption {
Behavior of $P(r,t)$ close to the origin $r=0$ for $\alpha=1/2$ and $d=3$.
We show $r P(r,t)$ versus $r$ for different times 
$t=0.02,0.2,2,200$
(dashed curves). Because $P(r,t)$ diverges when $r \to 0$ like $P(r,t)\sim 1/r$,
the figures shows that $rP(r,t) \to C_1(t)$.
Also shown (linear curves) the approximation
Eq.  
(\protect{\ref{eqExa02}}),
as expected when $\xi=r^2/t^{\alpha}<< 1$ the approximation
is in good agreement with the integral solution.
}
\label{fig2}
\end{figure}

More detailed behavior of $P(r,t)$ is presented in
Fig. \ref{fig2} where we show $r P(r,t)$ versus
$r$.  In Fig. \ref{fig2} we also exhibit linear curves based on
the asymptotic expansion Eq. (\ref{eqA15})
which predicts
\begin{equation}
r P(r,t)\sim C_1(t) -  r C_2(t)
\label{eqExa02}
\end{equation}
where
\begin{equation}
C_1(t)={1\over 4 \pi^{3/2}  t^{1/2}}
\label{eqExa03}
\end{equation}
and $C_2(t)=1/[4 \pi \Gamma(1/4) t^{3/4}]$.
Eq. (\ref{eqExa02}) is valid when $\xi=r^2/t^{1/2} << 1$ and 
as expected we see in Fig.
\ref{fig2}  that for this case the  numerical integral solution
 and the asymptotic solution
(\ref{eqExa02})
agree well.

%
%

%
%

\begin{figure}[htb]
\epsfxsize=20\baselineskip
\centerline{\vbox{
      \epsffile{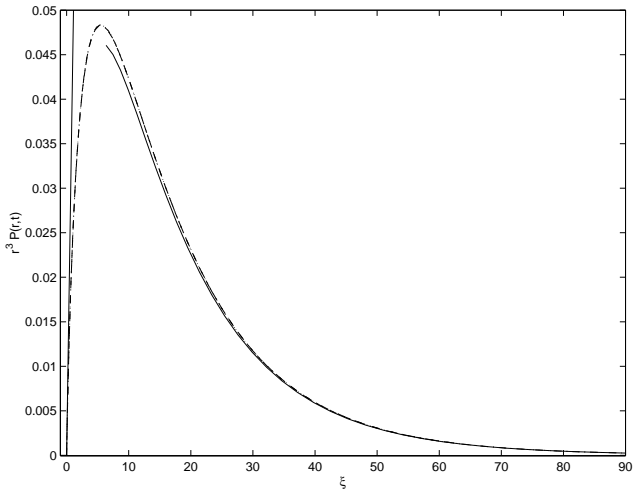}  }}
\caption {
Scaling behavior of $P(r,t)$ for $\alpha=1/2$ and $d=3$. We show $r^3 P(r,t)$ 
versus $\xi=r^2/t^{\alpha}$
for times $t=0.02$ (dotted), $t=2$ (dash dotted) and $t=200$  (dashed),
due to exact scaling of the solution one cannot distinguish between the three curves
in the figure. Also shown (solid curves) are the two asymptotic
behaviors valid for small and large values of the  scaling
variable $\xi$. For $\xi>>1$ we used Eq. (\protect{\ref{eq24}}) 
and for $\xi<<1$ we used the leading term in the expansion Eq.
(\protect{\ref{eqA15}}). 
}
\label{fig3}
\end{figure}

 Finally in Fig. \ref{fig3} we show our results in a scaling form,
similar to the way we presented the CTRW  results in Fig. \ref{figone}
which shows the CTRW solution for both short and long times.
We present $r^3 P(r,t)$ versus $\xi$ for different choices of 
time $t$. We observe collapse of all curves, both for shorter and longer times, 
onto one master curve.
We also show the asymptotic behaviors $\xi>>1$ and $\xi<<1$,
Eq. (\ref{eq24}) 
and  
Eq. (\ref{eqA15}) 
respectively. 
The numerical integral solution Eq. (\ref{eq05b})
agrees well with the asymptotic behaviors in the appropriate regimes.
Comparing Fig. (\ref{figone}) and Fig. (\ref{fig3}) we see that the solution
of the fractional diffusion equation approximates
the exact solution of the CTRW very well when $t \to \infty$
and $r\ne 0$.

%
%
\section{Fractional Fokker--Planck Equation}
\label{secFFPE}

 Let us consider the 
fractional Fokker--Planck equation (FFPE)
\cite{MBK} 
 describing the stochastic
evolution of a test particle under combined influence of
external force field $F(x)$ and a thermal heat bath. The equation
reads
\begin{equation}
\dot{P}\left( x , t \right) = K_{\alpha}\ _0 D_t^{ 1 - \alpha} 
\hat{L}_{fp} P\left(x,t\right)
\label{eqF01}
\end{equation}
and
\begin{equation}
\hat{L}_{fp} = - { \partial  \over \partial x} { F(x) \over k_b T } + { \partial^2 \over \partial x^2} 
\label{eqF02}
\end{equation}
is the Fokker--Planck operator. $K_{\alpha}$ is
a generalized diffusion constant and $T$ is temperature.
The fractional derivative $_0 D_t^{1-\alpha}$
was defined in Eq. 
(\ref{eq20aa}).
 We consider
the one dimensional case and extensions to higher dimensions
are straight forward. When $F(x)=0$ the FFPE Eq. (\ref{eqF01}) reduces
to the fractional diffusion equation
(\ref{eqa01}). When $\alpha = 1$ we recover the ordinary Fokker-Planck 
equation. The stationary solution of the FFPE is the Boltzmann distribution
and the equation is compatible with linear response theory \cite{MBK}. 

 In this section we find an integral solution of the FFPE.
We show that the solution is normalized and non negative,
an issue not discussed in \cite{MBK}. Without loss of generality we set    
$K_{\alpha}=1$. We consider initial condition $P(x,t=0) = \delta(x - x_0)$.

First let us show that the solution is normalized.
Integrating Eq. (\ref{eqF01}) with respect to $x$
and using the boundary conditions
\begin{equation}
P(x,t)|_{x=\pm \infty} = {\partial P(x,t) \over \partial x} |_{x = \pm \infty} = 0
\label{eqF03}
\end{equation}
we find
\begin{equation}
\int_{-\infty}^{\infty} \dot{P}(x,t) dx = 0,
\label{eqF04}
\end{equation}
meaning that normalization is conserved as it should.

 We write the solution of the FFPE Eq. (\ref{eqF01})
in terms of an integral of a  product of two functions
\begin{equation}
P(x,t) = \int_0^{\infty} ds \tilde{n}(s,t ) P_1 \left( x, s \right)
\label{eqF05}
\end{equation}
where
\begin{equation}
{\partial P_1(x,s) \over \partial s } = \hat{L}_{FP}P_1\left( x , s \right).
\label{eqF06}
\end{equation}
$P_1(x,s)$ is a normalized solution of the ordinary Fokker-Planck equation
with initial conditions $P_1(x,s=0)=\delta(x-x_0)$. Methods of 
solution of Eq. (\ref{eqF06}) are given in Risken's book \cite{Risken}.
In what follows we shall prove that $\tilde{n}(s,t)=n(s,t)$, 
defined in Eq.
(\ref{eq06}). 

 We use the Laplace transform of Eq. (\ref{eqF05}) and normalization
condition of $P(x,t)$ to show
\begin{equation}
\int_0^{\infty} \tilde{n}(s,u) ds = 1/u.
\label{eqF07}
\end{equation}
Hence $\tilde{n}(s,t)$ is normalized according to 
$\int_0^{\infty} \tilde{n}(s,t) ds=1$.
The Laplace transform of Eq. (\ref{eqF01}) reads
\begin{equation}
uP(x,u) - \delta(x-x_0) = u^{ 1 - \alpha} \hat{L}_{fp} P\left( x, u \right),
\label{eqF08}
\end{equation}
inserting Eq. (\ref{eqF05}) in Eq. (\ref{eqF08}) we find
$$ u \int_0^{\infty}\tilde{n}\left(s,u \right) 
P_1\left(x,s\right) ds - \delta(x - x_0)=$$
\begin{equation}
u^{1 - \alpha} \int_0^{\infty} \tilde{n}(s,u) 
\hat{L}_{fp} P_1\left( x , s\right) ds,
\label{eqF09}
\end{equation}
integrating by parts using Eq. (\ref{eqF06}) we find
$$ u \int_0^{\infty} \tilde{n}\left(s, u \right) 
P_1\left(x,s\right) ds - \delta(x - x_0)=$$
$$u^{1 - \alpha} 
\left[ \tilde{n}\left(\infty, u\right) P_1\left(x, s= \infty\right)
- \tilde{n}\left(0, u \right) P_1\left( x, 0 \right) \right]$$
\begin{equation}
- u^{1 - \alpha} \int_0^{\infty}\left[ { \partial \over \partial s} 
\tilde{n}\left( s,u \right)\right] P_1(x,s) ds.
\label{eqF10}
\end{equation}
From Eq. (\ref{eqF07}) $\tilde{n}(\infty,u)=0$ and since $P_1(x,0)= \delta(x-x_0)$ we may rewrite
Eq. (\ref{eqF10})
$$ \int_0^{\infty} \left\{ u \tilde{n}(s,u)
 +u^{1 - \alpha} \left[ { \partial \over \partial s} \tilde{n}\left(s, u \right)\right] \right\} P_1(x,s) ds=  
$$
\begin{equation}
\left[ 1 - u^{ 1 - \alpha} \tilde{n}(0,u)\right] \delta(x - x_0).
\label{eqF11}
\end{equation}
Eq. (\ref{eqF11}) is solved once both of its sides are equal
zero; therefore two conditions must be satisfied, the first
\begin{equation}
\tilde{n}(0,u) = u^{\alpha-1},
\label{eqF12}
\end{equation}
and the second
\begin{equation}
{ \partial \over \partial s} \tilde{n}\left(s,u\right)
= - u^{\alpha} \tilde{n}\left(s, u \right).
\label{eqF13}
\end{equation}
The solution of Eq. (\ref{eqF13}) with initial condition Eq. (\ref{eqF12})
is
\begin{equation}
\tilde{n}(s,u) = u^{ \alpha- 1} \exp\left( - u^{\alpha} s \right).
\label{eqF14}
\end{equation}
Thus $\tilde{n}(s,u) =n(s,u)$ 
found in the context of the solution of
CTRW Eq. (\ref{eq04aa}).

 We see that the integral solution of the FFPE has a similar structure
as the solutions of the fractional diffusion equation or
equivalently of the asymptotic CTRW. The solution
is  Eq.  
(\ref{eqF05})  with $n(s,t)$ defined in Eq. (\ref{eq06})
and $P_1(x,s)$ being the solution of corresponding 
ordinary Fokker-Planck equation.
It is easy to understand that the solution, $P(x,t)$
is normalized and non negative
because both $P_1(x,t)$ and $n(s,t)$ are normalized PDFs.

 In \cite{MBK} a different method of solution of the fractional
Fokker--Planck equation was investigated. It can be shown
that $P(x,t)$ can be expanded in an eigen function expansion,
which is similar to the standard eigen function expansion
of solution of ordinary  Fokker--Planck equation, however eigen
modes decay according to Mittag--Leffler relaxation instead of the
standard exponential decay of the modes in the integer
Fokker--Planck equation. It can be shown that the integral solution
investigated here is identical to the eigen function expansion
in \cite{MBK}. 

\subsection{Example 1, Biased Fractional Wiener Process}

 Consider the biased fractional diffusion process defined with a generalized
diffusion coefficient $K_{\alpha}$ 
and a uniform force field $F(x)=F>0$.
For this case the  mean displacement grows slower than linear with time
according to 
\begin{equation}
\langle x (t )
 \rangle = { F K_{\alpha} t^{\alpha} \over k_b T \Gamma(1 + \alpha)}.
\label{eqmean}
\end{equation}

 The well known solution of the 
ordinary Fokker-Planck equation is
\begin{equation}
P_1\left(x,s\right) = {1 \over 4 \pi s} \exp\left[ - {\left(x- \tilde{F} s\right)^2\over 4 s} \right].
\label{eqEx101}
\end{equation}
Eq. (\ref{eqEx101}) describes a biased Wiener process and $\tilde{F} = F/(k_b T)$.
The solution $P(x,t)$ for the fractional case is found using the transformation
Eq. (\ref{eqF05}). In Laplace $u$ space the solution is
$$ P(x,u) =  $$
\begin{equation}
{\tilde{F}  u^{ \alpha - 1} \tau^{\alpha} \over \sqrt{1 +  4 (u \tau)^{\alpha} }}
 \exp\left[ { \tilde{F} ( x - \sqrt{1+ 4 (u \tau)^{\alpha}}|x|)\over 2}\right]
\label{eqEx02}
\end{equation}
and $\tau^{\alpha} = 1/(\tilde{F} K_{\alpha})^2$. 
For $(\tau u )^{\alpha}<< 1$, corresponding to the long time behavior
of the solution $P(x,t)$, we find
\begin{equation}
P(x,u) \sim \left\{ 
\begin{array}{c|c}
\tilde{F} u^{ \alpha - 1} \tau^{\alpha} \exp\left( - \tilde{F} \tau^{\alpha} u^{\alpha} x\right)\
&   x> 0\\
\ & \ \\
\tilde{F} u^{ \alpha - 1} \tau^{\alpha} \exp\left( - \tilde{F} |x| - \tilde{F} |x| \tau^{\alpha} u^{\alpha} \right) &
 x \le 0.
\end{array}
\right.
\label{eqEx03}
\end{equation}
Since $\int_{- \infty}^0 P(x,u) \sim u^{\alpha - 1}$ 
and hence according to Tauberian theorem
$\int_{- \infty}^0 P(x,t) \sim 1/t^{\alpha}$ we have $P(x,t) =0$ for $x<0$ when $t \to \infty$.
Therefore, using the inverse Laplace transform of Eq. (\ref{eqEx03})
the asymptotic behavior of $P(x,t)$ is
\begin{equation}
\lim_{t \to \infty} P(x,t) =
\left\{
\begin{array}{c|c}
 { 1 \over \alpha A^{1/\alpha} } { t \over x^{1 + 1/\alpha}}
l_{\alpha}\left( { t \over A^{1/\alpha} x^{ 1 / \alpha} }\right)  &  0 < x \\
 \  & \ \\
0 &  0>x,
\end{array}
\right.
\label{eqEx04}
\end{equation}
and $A=\tilde{F} \tau^{\alpha}$.
Integrating Eq. (\ref{eqEx04}) we find the distribution function
\begin{equation}
\lim_{t\to  \infty} \int_{-\infty}^x P(x,t) dx= 1 - L_{\alpha}\left({t \over A^{ 1 /\alpha} x^{1/\alpha}}\right),
\label{eqEx05}
\end{equation}
valid for $x> 0$.
Eq. (\ref{eqEx05}) was
derived also in \cite{Marcin} based on the biased CTRW, thus as expected
the solution of the fractional Fokker-Planck equation converges to the solution
of the CTRW in the limit of large $t$.

On the origin one can use Tauberian theorem to show
\begin{equation}
P(0,t) \sim {A \over \Gamma( 1 - \alpha)} t^{ -\alpha}
\label{eqEx05a}
\end{equation}
valid for long times. For the case $F=0$ we have found
$P(0,t) \sim t^{ - \alpha / 2} $ so as expected the decay
on the origin is faster for the biased case since
particles are drifting away from the origin.

\begin{figure}[htb]
\epsfxsize=20\baselineskip
\centerline{\vbox{
      \epsffile{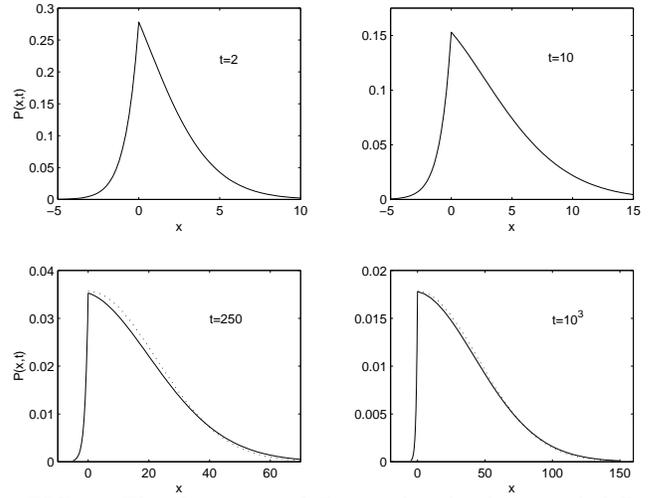}  }}
\caption {
The dynamics of $P(x,t)$ for the
fractional diffusion process in external field $F$ with $\alpha =1/2$. 
We show the solution
for four different times. The dotted curve is the asymptotic solution
valid for large $t$.
The maximum of $P(x,t)$ is
for all times on the initial condition  $x=0$.
We use $F=k_b T =K_{1/2}= 1$.
}
\label{figf4}
\end{figure}

  In Fig. \ref{figf4}
we present the solution for the case $\alpha = 1/2$, then 
$$ P(x,t) = $$ 
\begin{equation}
{1 \over \sqrt{ t K_{1/2}^2 \pi}} \int_0^{\infty} ds { 1\over \sqrt{4 \pi s}}
\exp\left[ -{ s^2 \over 4 K_{1/2}^2 t } - {(x - \tilde{F} s)^2\over 4s} \right]
\label{eqEx05b}
\end{equation}
which is evaluated numerically. For large times
we have
\begin{equation}
P(x,t)\sim { A \over \sqrt{\pi t} } \exp\left[ - {A^2 x^2 \over 4 t} \right]
\label{eqEx06}
\end{equation}
for $0 < x$. As seen in the figure the exact result exhibits a strong
sensitivity on initial condition and the maximum of $P(x,t)$ is located
on $x=0$. This is different than ordinary diffusion process in which the
maximum of $P(x,t)$ is on $\langle x(t) \rangle$. 
The curves in Fig. \ref{figf4} are similar to those observed
by Scher and Montroll \cite{SM} based on lattice simulation of CTRW and
also by Weissman et al \cite{WWH} who investigated biased CTRW
using an analytical
approach.
The FFPE solution presented here is much simpler
than the CTRW solution, still it captures all the important
features of 
the more complex
CTRW result.

%
%

\subsection{Example 2, the Fractional Ornstein-Uhlenbeck Process}
 
\begin{figure}[htb]
\epsfxsize=20\baselineskip
\centerline{\vbox{
      \epsffile{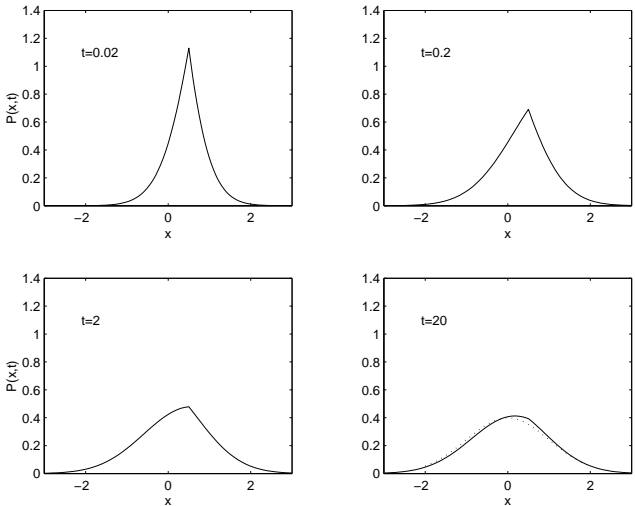}  }}
\caption {
The dynamics of $P(x,t)$ for the
fractional Ornstein--Uhlenbeck process and $\alpha =1/2$. We show the solution
for four different times. The dotted curve is the stationary solution,
i.e. the Boltzmann
distribution which for the Harmonic potential is Gaussian.
Notice the cusp on the initial condition $x=x_0=1/2$.
}
\label{figFUO}
\end{figure}

 We consider as a second 
example the fractional Ornstein--Uhlenbeck (OU) process,
namely the motion of the test particle in harmonic oscillator.
This case cannot be analyzed using the 
CTRW  in a direct way.
The CTRW formalism considers
only uniformly biased random walks and the fractional processes in
non-uniform fields are not uniformly biased.
We consider $\alpha=1/2$, $F(x)/k_b T= -x$ and use the well
known solution of the ordinary OU process \cite{Risken,Kampen}
\begin{equation}
P_1\left(x,s\right) = { 1 \over \sqrt{ 2 \pi \left( 1 - e^{ - 2 s} \right)}}
\exp\left[ - { \left( x-x_0 e^{ - s} \right)^2 \over 2 \left( 1 - e^{ - 2 s} \right)} \right].
\label{eqF15}
\end{equation}
The solution of the fractional OU process
is then found using numerical integration
of 
Eq. (\ref{eqF05}) using Eqs.  
(\ref{eqExam}) and (\ref{eqF15}). 
Our results are presented in Fig. \ref{figFUO}.
We have considered an initial condition $x_0=1/2$ and we observe a
strong dependence of the solution on the initial condition.
A cusp on $x=x_0$ is observed for all times $t$, thus the initial
condition has a strong influence on the solution. The
solution approaches the stationary Gaussian shape slowly in a power law
way and the solution
deviates from Gaussian for any finite time. Unlike the ordinary
Gaussian OU process,
the maximum of $P(x,t)$ is not on the average $\langle x(t) \rangle$ but
rather the maximum is for short times located  on the initial condition.

%
%

Like the ordinary OU process the fractional OU
process has a special role. The ordinary OU process describes two 
types of behaviors,
the first is an over damped  motion of a particle in harmonic potential, 
the second
is the velocity of a Brownian particle modeled by
the Langevin equation, the latter is the basis for the Kramers equation. 
In a similar way the fractional OU
describes an over damped and anomalous motion in Harmonic potential
considered in this section and in \cite{MBK}, it can also be used to model
the velocity of a particle exhibiting a L\'evy walk type of motion
\cite{BS}.
The fractional OU process
is the basis of the fractional Kramers equation introduced recently
by Barkai and Silbey \cite{BS}, 
this equation describes super 
diffusion while in
this work we have considered sub diffusion.

\section{Summary}
\label{secSUM}

 Fractional diffusion equation is an asymptotic equation which
predicts the behavior of the decoupled continuous time random
walk in the sub diffusive regime. The fractional Fokker--Planck
equation considers such a sub-diffusive motion in an external
force field and close to thermal equilibrium. In this work we have considered
an integral transformation which gives the solution of the fractional
Fokker--Planck equation in terms of solution of ordinary Fokker--Planck
equation. Solution of ordinary Fokker--Planck equation can be
found based on different analytical 
and numerical  methods \cite{Risken,Kampen,Hong1,Tai,Pollack}.
The integral transformation describes also the long time behavior of the CTRW
in dimension $d=1,2,3$. Thus the transformation maps Gaussian diffusion
onto fractional diffusion, and it can serve as a practical
tool for finding solution of certain fractional kinetic equations.

\section{Acknowledgments}

EB thanks A. I.  Saichev and G. M. Zaslavsky  for correspondence 
and J. Klafter, R. Metzler and G. Zumofen for discussions.

\section{Appendix A}

We rewrite Eq. (\ref{eqa02})
\begin{equation}
P\left({\bf k}, u \right)  \sim   u^{\alpha-1}\int_0^{\infty}
e^{-s\left( u^{\alpha} + {\bf k}^2\right)} ds,  
\label{eqa03a}
\end{equation}
the inverse Fourier transform of Eq. (\ref{eqa03a})
is 
\begin{equation}
P\left({\bf r}, u \right)  \sim   u^{\alpha-1}{\cal F}^{-1}\left\{
\int_0^{\infty} e^{-s\left( u^{\alpha} + {\bf k}^2\right)} ds \right\},
\label{eqa03b}
\end{equation}
and ${\cal F}^{-1}$ is the inverse Fourier transform. Changing the
order of integration over parameter $s$ and the operation
${\cal F}^{-1}$ (this is later justified by the identity
of moments of the integral solution and that found in \cite{schneider}) 
we find using Eq.
(\ref{eq04})
\begin{equation}
P\left({\bf r}, u \right)  \sim \int_0^{\infty}  n(s,u) { 1 \over \left( 4 \pi s\right)^{d/2} } \exp\left( - r^2/ 4s \right) ds.
\label{eqa03c}
\end{equation}
Applying the inverse Laplace transform to this equation,
changing the order of integrations over $s$ and the inverse Laplace
operation, we find
Eq. (\ref{eq05}). 

\section{Appendix B}

 We investigate the Cartesian moments of the CTRW.
These moments are 
$$
M\left( 2 m_1 , \cdots , 2 m_d\right) = 
$$
$$ \int_{-\infty}^{\infty} dx_1\cdots\int_{-\infty}^{\infty}d x_d x_1^{ 2 m_1} x_2^{m_2}\cdots x_d^{ 2 m_d} \times$$
$$ \sum_{s = 1}^{\infty}\left\{ L_{\alpha}\left[{ t \over s^{1/\alpha}}\right]
- L_{\alpha}\left[ { t \over \left( s + 1\right)^{ 1/\alpha}}\right]\right\}\times$$
\begin{equation}
{1 \over \left(4 \pi s \right)^{d/2}}\exp\left( - { x_1^2+ \cdots x_d^2\over 4 s}\right).
\label{eqWW01}
\end{equation}
Changing order of integration and summation, 
we use the identity 
$$
\int_{-\infty}^{\infty} dx_1\cdots \int_{-\infty}^{\infty}dx_d 
{1 \over \left(4 \pi s \right)^{d/2}}\exp\left( - { x_1^2+ \cdots x_d^2\over 4 s}\right)\times$$
\begin{equation}
x_1^{2 m_1} \cdots x_d^{ 2 m_d}=
C_{m,d} s^m
\label{eqB02aa}
\end{equation}
were
$C_{m,d}$ is defined in Eq. (\ref{eqffff1})
Inserting Eq. (\ref{eqB02aa}) in Eq.
(\ref{eqWW01})
 we find Eq. 
(\ref{eqffff}).

The Laplace transform of Eq. (\ref{eqffff}) is
\begin{equation}
M\left( 2 m_1 , \cdots, 2 m_d\right)= 
C_{m,d}{1 - e^{ - u^{\alpha}} \over u} \sum_{s=1}^{\infty} e^{ - s u^{\alpha}} s^m.
\label{eqfffft}
\end{equation}
We use
$$\sum_{s=1}^{\infty} e^{ - s u^{\alpha}} s^m =
(-1)^m \left( { d \over dx} \right)^m \sum_{s = 1}^{\infty} e^{ - x s} |_{x=u^{\alpha}}=$$
\begin{equation}
\left( - 1 \right)^m \left( { d \over dx} \right)^m{ e^{- x} \over 1 - e^{ - x} } |_{x=u^{\alpha}},
\label{equuu}
\end{equation}
and for small $u^{\alpha}$ we find
\begin{equation}
M\left( 2 m_1 , \cdots, 2 m_d\right)\sim C_{m,d}\Gamma( m + 1) u^{ - 1 - \alpha m}.
\label{equuau}
\end{equation}
Applying Tauberian theorem [i.e., inverting Eq. (\ref{equuau})]
  we find Eq. (\ref{eqddd}).

\section{Appendix C}

 We consider the moments 
$$
M\left( 2 m_1 , \cdots , 2 m_d\right) = 
$$
\begin{equation}
\int_{-\infty}^{\infty} dx_1\cdots\int_{-\infty}^{\infty}d x_d
\left[ \int_0^\infty n\left(s,t\right) G\left({\bf r},s\right)\right]
x_1^{2 m_1} \cdots x_d^{ 2 m_d},
\label{eqB01}
\end{equation}
changing the order of integration over $s$ and the $d$ dimensional
integration, 
using  
Eq. (\ref{eqB02aa}) we find
$$
M\left( 2 m_1 , \cdots , 2 m_d\right) = 
$$
\begin{equation}
C_{m,d}\int_0^{\infty} s^m n(s,t) ds.
\label{eqB03}
\end{equation}
were $C_{m,d}$ is defined in Eq. (\ref{eqffff1})
and $m=\sum_{i=1}^d m_i$,
Using Eq. 
(\ref{eq06}) the integral in Eq.
(\ref{eqB03})
is 
$$ I_m(t) \equiv \int_0^{ \infty} s^m n\left(s, t \right) ds = $$
\begin{equation}
{ 1 \over \alpha} \int_0^{ \infty} { 1 \over t^{ \alpha} }   
\left({ t \over s^{ 1 / \alpha} } \right)^{\alpha + 1} l_{\alpha} \left( t / s^{ 1/\alpha}\right) s^m ds.
\label{eqB03a}
\end{equation}
Notice that $I_m(t)$ is the $m$ moment of $n(s,t)$.
Changing the integration variable, in Eq. (\ref{eqB03a}),
 according to
$y=t/s^{ 1/ \alpha} $  we
find
\begin{equation}
I_m(t) = t^{ \alpha m } \int_0^{ \infty} y^{ - \alpha m } l_{ \alpha}\left( y \right) dy,
\label{eqB04}
\end{equation}
the calculation of the negative moment in Eq. (\ref{eqB04})
is straight forward using Laplace transform technique \cite{Weiss1},
we find
\begin{equation}
I_m(t) = { t^{ \alpha m } \over \Gamma\left( \alpha m \right) } 
\int_0^{ \infty} u^{ \alpha m - 1} l_{\alpha}\left( u \right) du
\label{eqB05}
\end{equation}
with $l_{ \alpha} \left( u \right) = \exp( - u^{ \alpha} )$ being the
Laplace transform of the one sided L\'evy density. Therefor we find
\begin{equation}
I_m(t) = { \Gamma\left( 1 + m \right) \over \Gamma\left(1 +  \alpha m \right)}
t^{ \alpha m }.  
\label{eqB06}
\end{equation}
Inserting Eq. (\ref{eqB06}) in 
Eq. (\ref{eqB03}) we find
the result Eq. (\ref{eqIS07}).
Notice that when $\alpha = 1$, $I_m(t)= t^m$, this is expected since
for this case $n(s,t) = \delta(s-t)$.

\section{Appendix D}

Integral formulas involving the product of two H Fox functions
are a helpful tool with which H functions can be represented
in terms of known functions. According to \cite{sri}
equation $(5.1.2)$
$$ \int_0^{\infty} dx \left\{\right. x^{\eta - 1}
H_{p,q}^{m,n}\left[ z x^{ - \sigma}\left|
\right.
\begin{array}{l}
\left(a_j,\alpha_j\right)_{1,p}\\
\left(b_j, \beta_j \right)_{1,q}
\end{array}
\right]
$$
$$ \times H_{P,Q}^{M,N}\left[ s x\left|
\right.
\begin{array}{l}
\left(c_j,\gamma_j\right)_{1,P}\\
\left(d_j, \delta_j \right)_{1,Q}
\end{array}
\right]
\left. \right\}
=
$$
$$ s^{ - \eta} H_{p + P , q + Q}^{m + M , n + N} 
\left[ \right. z x^{\sigma} \left| \right. $$
\begin{equation}
\begin{array}{l}
\left(a_j,\alpha_j\right)_{1,n},
\left(c_j + \eta \gamma_j, \sigma \gamma_j\right)_{1, P},
\left(a_j,\alpha_j\right)_{n+1, p }\\
\left(b_j, \beta_j \right)_{1,m},
\left(d_j+\eta \delta_j, \sigma \delta_j \right)_{1,Q},
\left(b_j,\beta \right)_{m+1,q},
\end{array}
\left. \right],
\label{IntFor}
\end{equation}
provided that seven conditions are satisfied, $\sigma>0$,
\begin{equation}
\delta=\sum_{j = 1}^q \beta_j - \sum_{j = 1}^p \alpha_j> 0,
\label{eqAppCC01}
\end{equation}
\begin{equation}
\delta'=\sum_{j = 1}^Q \delta_j - \sum_{j = 1}^P \gamma_j> 0,
\label{eqAppCC02}
\end{equation}
\begin{equation}
A=\sum_{j=1}^n \alpha_j - \sum_{j=n+1}^p \alpha_j+\sum_{j=1}^m\beta_j-\sum_{j=m+1}^q\beta_j>0,
\label{eqAppCC03}
\end{equation}
\begin{equation}
A'=\sum_{j=1}^M \delta_j - \sum_{j=M+1}^Q \delta_j+\sum_{j=1}^N\gamma_j-\sum_{j=N+1}^P\gamma_j>0,
\label{eqAppCC04}
\end{equation}
\begin{equation}
\sigma
\begin{array}{c}
\ \\
 \mbox{max}\\
1\le j \le n
\end{array}
\left[\left(a_j -1\right)/\alpha_j\right] -
\begin{array}{c}
\ \\
 \mbox{min}\\
1\le j \le M
\end{array}
\left[d_j/\delta_j\right] < \eta
\label{eqAppCC04a}
\end{equation}
and
\begin{equation}
\eta< \sigma
\begin{array}{c}
\ \\
\mbox{min}\\
1\le j \le m
\end{array}
\left[b_j/\beta_j \right]-
\begin{array}{c}
\ \\
\mbox{max}\\
1\le j \le N
\end{array}
\left[\left(c_j-1\right)/\gamma_j \right].
\label{eqAppCC05}
\end{equation}
We have considered the case when all  parameters are real, for the case when
parameters may become complex see \cite{sri,mathai}.

 We express the integral solution Eq.
(\ref{eq05b}),
in terms of an integral of a  product of
two Fox H functions, after some rearrangements
and with the use of Eq. (\ref{eqLeSc}) and  the following identity
\begin{equation}
G\left( s , r \right) = \pi^{ - d/2} r^{ - d} H_{0,1}^{1,0}\left[ \right.
{ r^2 \over 4 s} \left| \right.
\begin{array}{l}
-- \\
\left( d/2, 1 \right)
\end{array}
\left]\right.
\label{eqAppCC06}
\end{equation}
we find
$$ P\left( r, t\right)= \int_{0}^{\infty} n(s,t) G\left( r,s\right)ds= $$
$$ \alpha^{-2} \pi^{-d/2} r^{ - d} \left\{ \right. $$
$$ \int_0^{\infty} dx x^{ - 1 - 1/\alpha} H_{1,1}^{1,0}\left[ \right. 
x^{ -1 /\alpha} \left| \right.
\begin{array}{l}
\left( - 1, 1 \right)\\
\left( - 1/\alpha, 1/\alpha\right)
\end{array}
\left. \right] \times $$
\begin{equation}
H_{0, 1}^{1, 0 } \left[ {r^2 \over 4 t^{\alpha} } x \left| \right.
\begin{array}{l}
-- \\
\left( d/2, 1 \right)
\end{array}
\right].
\label{eqAppCC06a}
\end{equation}
Using  Eq. (\ref{IntFor}) one can hope to prove that the integral solution
Eq. (\ref{eq05b}) and the Fox Function solution  Eq. (\ref{eq23})
are identical, 
provided of course that all conditions Eqs. (\ref{eqAppCC01}-\ref{eqAppCC05}) 
are satisfied. Unfortunately the integral identity Eq. (\ref{IntFor}) cannot be used
for this aim because condition 
Eq. (\ref{eqAppCC05}) is not satisfied, inserting the fractional 
parameters in Eq. (\ref{eqAppCC05}) we find 
\begin{equation}
-1/\alpha < - 1/\alpha
\label{eqAppC007}
\end{equation}
which shows that the conditions are not fulfilled.

 It is worthwhile mentioning that all the other conditions (\ref{eqAppCC01}-\ref{eqAppCC04a})
are fulfilled and that it is possible to prove that 
the integral solution and the Fox function solution are identical if we 
replace $<$ with $\le$ in Eq. (\ref{eqAppCC05}). 
Using Eq. 
(\ref{IntFor}), the identity
$$ \alpha^{-1} H_{1,2}^{2,0} \left[ x^{1/\alpha} \left| \right. 
\begin{array}{l}
\left( 0, 1 \right)\\
\left( 0 , 1/ \alpha\right), \left(d/2, 1/\alpha\right)
\end{array}
\right]=$$
\begin{equation}
H_{1,2}^{2,0} \left[ x^{1/\alpha} \left| \right. 
\begin{array}{l}
\left( 1, 1 \right)\\
\left( d/ 2 , 1/ \alpha\right), \left(1, 1/\alpha\right)
\end{array}
\right]=
\label{eqACC08}
\end{equation}
which can be proven based upon the definition of the Fox function, 
and  \cite{sri,mathai}
$$ z^{\sigma} H_{p,q}^{m,n} \left[ z \left| \right.
\begin{array}{l}
\left( a_j,\alpha_j\right)_{1,p}\\
\left(b_j,\beta_j\right)_{1,q}
\end{array}
\right] =$$
\begin{equation}
H_{p,q}^{m,n} \left[ z \left| \right.
\begin{array}{l}
\left( a_j,\sigma \alpha_j\right)_{1,p}\\
\left(b_j,\sigma \beta_j\right)_{1,q}
\end{array}
\right] 
\label{eqAppCC08}
\end{equation}
we find
$$ P\left(r, t\right)= \int_{0}^{\infty} n(s,t) G\left(r,s\right)ds= $$
$$  \alpha^{-1} \pi^{-d/2} r^{ - d} \times $$
\begin{equation}
H_{12}^{20}
\left( 2^{- 2 /\alpha} r^{2 /\alpha} t^{-1} |
\begin{array}{l}
\left( 1 , 1 \right) \\
\left( d/2, 1/\alpha),(1,1/\alpha\right)
\end{array}
\right)
\label{edq23mir}
\end{equation}
which is the Fox function  solution of Eq. (\ref{eq23}).
We emphasize that this equation was not proven in this Appendix
because condition in Eq. 
(\ref{eqAppCC05}) was not satisfied. 

\section{Appendix E}

 The Fox function is represented as
\begin{equation}
H^{m,n}_{p,q}\left[ x \left|
\right. 
\begin{array}{l}
\left(a_1,\alpha_1\right)\cdots\left(a_p , \alpha_p\right)\\
\left(b_1, \beta_1 \right) \cdots\left(b_q,\beta_q\right)
\end{array}
\right],
\label{eqA01}
\end{equation}
and for our choice of parameters
$$ m=2,n=0,p=1,q=2$$
$$ a_1=1, \alpha_1=1$$
\begin{equation}
b_1=d/2, \ \ \beta_1=1/\alpha , \ \ b_2=1, \ \ \beta_2=1/\alpha
\label{eqA02}
\end{equation}
The asymptotic expansion of the H Fox function, for $0<x<<1$, 
is defined when two conditions are satisfied \cite{sri,mathai}.
The first is
\begin{equation}
\delta=\sum_{j=1}^q\beta_j - \sum_{j=1}^p \alpha_j > 0, 
\label{eqA03}
\end{equation}
and for the case $\delta =  0$ see \cite{sri,mathai}.
For our case, defined by the parameters in Eq.
(\ref{eqA02}), $\delta=2/\alpha-1> 0$ when $0<\alpha < 1$.
The second condition is 
\begin{equation}
\beta_h\left( b_j + \Lambda\right) \ne \beta_j\left(b_h +  k\right)
\label{eqA04}
\end{equation}
for 
\begin{equation}
j \ne h; \ \ j=h=1,\cdots m; \ \ \Lambda,  k=0,1,2,\cdots .
\label{eqA05}
\end{equation}
Using Eq. (\ref{eqA02}) condition Eq. (\ref{eqA05}) reads
\begin{equation}
{1 \over \alpha} \left( 1 + \Lambda\right) \ne 
{1\over \alpha} \left( d/2 + k \right)
\label{eqA06}
\end{equation}
therefore the condition is satisfied
for dimensions $d=1$ and $d=3$ but not for $d=2$.
When conditions are satisfied 
$$ H^{m,n}_{p,q}\left[ x \left|
\right. 
\begin{array}{l}
\left(a_1,\alpha_1\right)\cdots\left(a_p , \alpha_p\right)\\
\left(b_1, \beta_1 \right) \cdots\left(b_q,\beta_q\right)
\end{array}
\right]=$$
$$ \sum_{h=1}^m \sum_{k=0}^{\infty} \Pi_{j=1,j\ne h}^m \Gamma\left( b_j - \beta_j \xi_{h,k}\right) \Pi_{j=1}^n \Gamma\left(1- a_j + \alpha_j \xi_{h, k} \right)\times$$
$$ \left( - 1 \right)^k \left(x\right)^{\xi_{h,k}}
\left[ \Pi_{j=m+1}^q \Gamma\left( 1 - b_j + \beta_j \xi_{h,k}\right) \right.$$
\begin{equation}
\left. \Pi_{j=n+1}^p \Gamma\left( a_j - \alpha_j \xi_{h,k} \right) k! \beta_h \right]^{-1}
\label{eqA07}
\end{equation}
where
\begin{equation}
\xi_{h,k} = \left( b_h + k \right)/ \beta_h .
\label{eqA08}
\end{equation}
Using Eq. (\ref{eqA07}) we find
$$
H^{2,0}_{1,2}\left[ x \left|
\right. 
\begin{array}{l}
\left(1,1\right)\\
\left(d/2, 1/\alpha\right),\left(1,1/\alpha\right)
\end{array}
\right]=
$$
$$
\alpha\left\{ \right. \sum_{k=0}^{\infty} { \Gamma\left( 1 - d/2 - k \right) 
\left( - 1 \right)^k x^{ \alpha(d/2 + k )} \over
\Gamma\left( 1 - \alpha d / 2 - \alpha k \right) k!}
$$
\begin{equation}
+ \sum_{k=0}^{\infty} { \Gamma\left( d/2 - 1 - k \right) \left( -1 \right)^k
x^{\alpha( 1 + k ) } \over
\Gamma\left( 1 - \alpha - \alpha k \right) k!}
\left. \right\} .
\label{eqA09}
\end{equation}
%
Using some simple manipulations we find
our results Eqs.
(\ref{eqA14}) and  
(\ref{eqA15}).


\begin{thebibliography}{99}

%
\bibitem{Weiss7}  E. W. Montroll and G. H. Weiss,
                 {\em J. Math. Phys.} {\bf 6}, 167 (1965)

%
\bibitem{Weiss1} G. H. Weiss, {\em  Aspects and Applications of the Random Walk}
               North Holland (Amsterdam -- New York -- Oxford, 1994)

\bibitem{shlesinger} M. F. Shlesinger, {\em J. Stat. Phys}, {\bf 10}, 421 (1974). 

\bibitem{SM} H. Scher and E. Montroll, {\em Phys. Rev. B}  {\bf 12}, 2455 (1975)

\bibitem{Bouch} J.--P. Bouchaud and A. Georges, {\em Phys. Rep.}
{\bf 195}, 127 (1990)

%
\bibitem{KBS} J. Klafter, A. Blumen and M. F. Shlesinger, {\em Phys. Rev.
A} {\bf 35}, 3081 (1987).

%
\bibitem{SKW} M. F.  Shlesinger, B. West and J. Klafter, {\em Phys. Rev.
Lett.},
                 {\bf 58}, 1100 (1987).


\bibitem{Klafter1} J. Klafter, M. F. Shlesinger and G. Zumofen, {\em Phys. Today
} {\bf 49} (2) 33 (1996).

\bibitem{barkai8} E. Barkai and J. Klafter, 
Lecture Notes in Physics, S. Benkadda and G. M. Zaslavsky
Ed. Chaos, Kinetics and Non-linear Dynamics in Fluids and Plasmas
(Springer-Verlag, Berlin 1998).

\bibitem{barkai10} E. Barkai, J. Klafter and V. Fleurov, {\em Phys. Rev. E}

\bibitem{Antony} A. Torcini and M. Antony,
                   {\em Phys. Rev. E} {\bf 57}, R6233 (1998)

\bibitem{latora} V. Latora, P.  Rapisarda and  S. Ruffo,
{\em Phys. Rev. Lett.} {\bf 83} 2104 (1999)

%
\bibitem{Silbey} J. Klafter and R. Silbey,
              {\em  Phys. Rev. Lett.} {\bf  44}, 55  (1980).


%
\bibitem{Levitz} P. Levitz, {\em Europhys. Lett.}, {\bf 39}, 593 (1997).

%
\bibitem{Zumofen} G. Zumofen and J. Klafter, {\em Phys. Rev. E} 
{\bf 47}, 851 (1993).


\bibitem{Swin} T. H. Solomon, E. R. Weeks and H. L. Swinney,
                   {\em Phys. Rev. Lett.} {\bf 71}, 23 (1995).

\bibitem{barkai11} E. Barkai and J.  Klafter, {\em Phys. Rev. Lett.} {\bf 79}, 2245,  (1997)

%


\bibitem{kuz} D. Kusnezov, A. Buglac and G. D. Dang,
{\em Phys. Rev. Lett.}  {\bf 82}, 1136 (1999)

\bibitem{fogedby1}   
H. C. Fogedby, {\em  Phys. Rev. Lett.} {\bf 73}, 2517 (1994);
Phys. Rev. E {\bf 58}, 1690 (1998); S. Jespersen, R. Metzler and H. C.
Fogedby, {\em Phys. Rev. E} {\bf 59},
2736 (1999)

\bibitem{zaslavsky}
G. M. Zaslavsky, M. Edelman and B. A. Niyazov,
{\em Chaos} {\bf 7}, 159 (1997)

\bibitem{Mainardi} F. Mainardi, {\em Appl. Math. Lett.}
{\bf 9} (6) 23 (1996)

\bibitem{Kol} K. M. Kolwankar and A. D. Gangal, {\em Phys. Rev. Lett.}
{\bf 80}, 214 (1998)

\bibitem{Rocco} Grigolini P, Rocco A, West B. J., {\em Phys. Rev. E.}
{\bf 59} 2603 (1999), Rocco A, West B. J., {\em Physica A},
{\bf 265} 535 (1999)

\bibitem{Hui} T. Huillet, {\em J. Phys. A}  {\bf 32} 7225 (1999)


\bibitem{oldham} K. B. Oldham and J. Spanier, {\em The
fractional calculus} Academic Press, (New York) 1974.

\bibitem{schneider} W. R. Schneider and  W. Wyss, {\em  J. Math. Phys.}
{\bf 30}, 134 (1989)

\bibitem{Hilfer} R. Hilfer and L. Anton, {\em Phys. Rev. E.} {\bf 51},
R848 (1995)

\bibitem{compte} A. Compte, {\em  Phys. Rev. E} {\bf   53},  4191 (1996)

\bibitem{barkai9} E. Barkai, R. Metzler and J. Klafter, {\em Phys. Rev. E.}
{\bf 61} 132 (2000)

\bibitem{MBK} R. Metzler, E. Barkai and J. Klafter, {\em  Phys. Rev. Lett.} {\bf 82},
             3563 (1999) 

\bibitem{Gurt} J. Klafter and G. Zumofen, {\em J. Phys. Chem.}
{\bf 98}, 7366 (1994)

\bibitem{saichev} A. I. Saichev and M. Zaslavsky, {\em  Chaos} {\bf 7} 4 1997

\bibitem{BS} E. Barkai, R. Silbey , {\em J. Phys. Chem.}

\bibitem{Tunaley} J. K. E. Tunaley, {\em J.  Stat. Phys.} {\bf 11}, 397 (1974). 

\bibitem{BHW} R. Ball, S. Havlin and G. H. Weiss, {\em J. Phys. A.}
{\bf 20}, 4055 (1987)

\bibitem{WWH} H. Weissman, G. H. Weiss, S. Havlin, {\em J. Stat. Phys.}
{\bf 57}, 301 (1989)

\bibitem{Marcin} M. Kotulski, {\em  J. Stat. Phys.} {\bf 81}, 777 (1995)

\bibitem{Feller} W. Feller,{\em  An introduction to probability Theory and
Its Applications} Vol. 2 (John Wiley and Sons 1970).

\bibitem{Sch1} W. R. Schneider  in  {\it Stochastic Processes in
Classical and Quantum Systems} Eds. S. Albeverio, G. Casatti and D. Merlini
(Lecture Notes in Physics, Springer, Berlin, 1986)

\bibitem{remark} Usually the choice $\psi(u)=1/(1 + u^{\alpha})$ is made
and it is assumed that the asymptotic results of the CTRW will not
differ for other choices behaving like $\psi(u)= 1 - u^{\alpha}+ \cdots$.

\bibitem{sri} H. M. Srivastava, K. C. Gupta and S. P. Goyal,
{\it The $H$--functions of one and two variables
with applications\/} (South Asian Publishers, New Delhi, 1982)

\bibitem{mathai} A. M. Mathai and R. K. Saxena, {\it The
$H$--function with Applications in Statistics and Other
Disciplines\/} (Wiley Eastern Ltd, New Delhi, 1978)


\bibitem{Balakrishnan} V. Balakrishnan, {\em Physica A} {\bf 132} 569 (1985)

\bibitem{Wyss}  W. Wyss, {\em J. Math. Phys.}
{\bf 27}, 2782 (1986)

\bibitem{SaichevPC}
A. I. Saichev private communication

\bibitem{Gl} W. G. Gl$\ddot{o}$ckle and T. F. Nonnenmacher, Macromolecules, {\bf 24}
6426 (1991)

\bibitem{Risken} H. Risken, {\it The Fokker--Planck
equation} (Springer, Berlin, 1989)

\bibitem{Kampen} N.G. van Kampen {\em Stochastic Processes in Physics and
               Chemistry} North Holland (Amsterdam -- New York -- Oxford, 1981)

\bibitem{Hong1} K. M. Hong and J. Noolandi, {\em J. Chem. Phys.} {\bf 68(11)}
5163 (1978)

\bibitem{Tai} H. Sano and M. Tachiya, {\em J. Chem. Phys.} {\bf 71 (3)} 1279
(1979)

\bibitem{Pollack} E. Pollak, {\em J. Chem. Phys.} {\bf 99 } 1344 (1993)

\end{thebibliography}
\end{document}